\documentclass{aip-cp}

\usepackage[numbers]{natbib}
\usepackage{rotating}
\usepackage{graphicx}
\usepackage{mathrsfs}
\usepackage{amssymb}
\usepackage{graphicx}
\usepackage[normalem]{ulem}
\usepackage[dvips]{color}
\usepackage{bm}
\usepackage{longtable}
\usepackage{times}

\renewcommand\sout{\bgroup \color{red} \ULdepth=-.5ex \ULset}

\renewcommand{\rm}[1]{\textrm{#1}}

\begin{document}

\title{Mean-field potential effects in the cumulants of baryons from central Au+Au collision at $E_{lab}$= 1.23 GeV$/$nucleon}
\author[aff1]{Yongjia Wang\corref{cor1}}
\corresp[cor1]{Corresponding author and speaker: wangyongjia@zjhu.edu.cn}
\author[aff1,aff2]{Yunxiao Ye}
\author[aff1,aff3]{Qingfeng Li}
\affil[aff1]{School of Science, Huzhou University, Huzhou 313000, China}
\affil[aff2]{Department of Physics, Zhejiang University, Hangzhou 310027, China}
\affil[aff3]{Institute of Modern Physics, Chinese Academy of Sciences, Lanzhou 730000, China}
\maketitle

\begin{abstract}
The cumulants of baryon multiplicity distribution in relativistic heavy-ion collisions (HICs) have attracted considerable attention recently. It has been conjectured that they may serve as a promising observable to detect the critical end point in the QCD phase diagram, while the cumulants in HICs at intermediate energies have not been widely studied to date. How to interpret the cumulants data at intermediate energies and compare with the data at relativistic energies is now being actively discussed. Both meam-field potential and clustering are highly important to HICs at intermediate energies. In this talk, we discuss these effects on the cumulant ratios of baryon number distributions in Au+Au collisions at beam energies of 1.23 GeV$/$nucleon which have been currently performed by the HADES Collaboration at GSI. Within the newest version of the ultrarelativistic quantum molecular dynamics (UrQMD) model, calculations with different mean field potentials as well as without mean field potential are performed. It is found that the mean field potential enhances fluctuations in the momentum space during the expanding stage, especially in a small rapidity acceptance window. The enhancement of cumulant ratios for free protons is suppressed compared with that for all baryons.
\end{abstract}

\section{Cumulants in relativistic heavy-ion collisions }\label{sec2}

 Exploring the quantum chromodynamic (QCD) phase diagram is one of the most important motivations for the study of relativistic heavy-ion collisions (HICs) physics.
 It is known from Lattice QCD methods that at zero baryon chemical potential (at the highest beam energies), phase transition from hadronic matter to the quark-gluon plasma (QGP)
 is a smooth crossover, while various effective theoretical models predict that this phase transition is expected to be a first-order one in the region of large baryon chemical potential (at lower energies).
 Thus, the QCD critical point should exist at a certain temperature and baryon chemical potential. To find the location of this critical end point in the temperature and baryon chemical potential plane has been
 one of the prime goals for studying of HICs physics, such as the beam energy scan (BES) program at the relativistic heavy-ion collider (RHIC), the future accelerator facilities FAIR in Germany and NICA in Russia. The basic idea is to search for the non-monotonic behavior of observables (such as fluctuations) as a function of the colliding energy. To determine the critical point from HICs, the fluctuations of conserved quantities, such as baryon, electric charge, and strangeness number, which are expected to be sensitive to the QCD phase transition and the critical point,
 have been considered as a promising observable\cite{Adams:2012th,Luo:2017faz,Gupta:2011wh}.

During the first phase of the BES program at RHIC (from the year 2010 to 2014), the cumulants (variance, skewness, and kurtosis, up to the fourth order) of net-proton (net-baryon), net-charges, and net-kaon number distribution in Au+Au collisions at energies of $\sqrt{s_{NN}}$ = 7.7, 11.5, 14.5, 19.6, 27, 39, 62.4 and 200 GeV have been measured\cite{Adare:2015aqk,Adamczyk:2014fia,Adamczyk:2013dal,Luo:2015ewa}. The almost flat energy dependence of cumulants for net-charge and net-kaon distributions was observed within current statistics, while the kurtosis of net-proton distributions in the most central (0-5\%) Au+Au collisions shows a non-monotonic behavior as a function of energy and a large deviation from unity at $\sqrt{s_{NN}}$=7.7 GeV. This result implies the critical end point in the QCD phase diagram may be reached or closely approached in the most central Au+Au collisions at $\sqrt{s_{NN}}$=7.7 GeV. While various effects, such as system volume fluctuations,
efficiency corrections, baryon clustering, global charge conservation, resonance decays (making correlations between positive and negative charges), may also influence the cumulants to some extent. To reach higher statistical precisions and to achieve a better understanding of the non-monotonic behavior for the kurtosis of the net-proton distributions, the second phase of BES program and the STAR Fixed-Target program have been proposed to measure cumulants at lower energies\cite{Meehan:2016qon}. In the near future, fixed target collisions at $\sqrt{s_{NN}}=2.7 \sim 4.9$ GeV in the Compressed Baryonic Matter (CBM) experiment on FAIR at GSI\cite{Ablyazimov:2017guv}, the J-PARC heavy-ion project in Japan\cite{Sako:2016edz}, as well as the NICA project in Russia\cite{Sissakian:2009zza} may also contribute to a deeper insight of the QCD phase diagram.

\section{The UrQMD model and cumulants in central Au+Au collision at intermediate energies}
At intermediate energies (beam energy around 1 GeV$/$nucleon), one can not expect phase transition from hadronic matter to the quark-gluon plasma taking place, but the nuclear liquid-gas phase transition may occur. Recently, the HADES collaboration showed new preliminary results for the cumulants of proton multiplicity distributions in Au+Au collisions at beam energy of 1.23 GeV$/$nucleon\cite{Caines:2017vvs}. How do the cumulants behave at intermediate energies and how to compare the cumulants at high energies to that at low energies is now being actively discussed. At intermediate energies, dense nuclear matter (twice to three times saturation density) can be created and a large
fraction of the protons and neutrons is bound in fragments. Both the collective flow and the nuclear stopping reach their maximum here. Thus, the cumulants for particle multiplicity distribution at intermediate energies are very complicated to evaluate.

The ultrarelativistic quantum molecular dynamics (UrQMD) model is a microscopic many-body transport approach in which each hadron is represented by Gaussian wave packet in phase space.
The time evolution of the centroids ($\textbf{r}_i$ and $\textbf{p}_i$) of the Gaussians obey Hamilton's equations,
\begin{eqnarray}
\dot{\textbf{r}}_{i}=\frac{\partial  \langle H  \rangle}{\partial\textbf{ p}_{i}},
\dot{\textbf{p}}_{i}=-\frac{\partial  \langle H \rangle}{\partial \textbf{r}_{i}}.
\end{eqnarray}
Here {\it $\langle H \rangle$} is the total Hamiltonian function of the system, it consists of the kinetic energy of the particles and the effective interaction potential energy. The UrQMD model has been widely and successfully used to study pp, pA, and AA collisions within a large energy range from the Fermi energy up to the CERN Large Hadron Collider (LHC) energies\cite{Bass:1998ca,Bleicher:1999xi}. It has been found that with a proper set of the in-medium nucleon-nucleon cross section, Pauli blocking, as well as the clustering process, the collective flow, stopping observable, as well as fragment mass distribution in intermediate energy HICs can be reproduced fairly well\cite{Li:2018wpv,du,wyj1417,Wang:2018hsw}. For studying HICs at intermediate energies, the following density and momentum dependent potential has been widely used\cite{Aichelin:1991xy,Hartnack:1997ez,Li:2005gfa},
\begin{equation}\label{eq2}
U=\alpha (\frac{\rho}{\rho_0})+\beta (\frac{\rho}{\rho_0})^{\gamma} + t_{md} \ln^2[1+a_{md}(\textbf{p}_{i}-\textbf{p}_{j})^2]\frac{\rho}{\rho_0}.
\end{equation}
Here $\alpha$, $\beta$, $\gamma$, $t_{md}$, and $a_{md}$ are parameters which can be adjusted to yield different nuclear equation of state (EoS) for isospin symmetric nuclear matter. In order to study the influence of mean field potential on the cumulants, the so-called soft and momentum dependent (SM), and hard and momentum dependent (HM) nuclear EoS, as well as the hard and without momentum dependent (H) EoS are chosen. In addition, the UrQMD model without any mean field potential (cascade mode) is also chosen. The set of parameters are displayed in Table I\cite{Li:2005gfa}.
\begin{table}[htbp]
\centering
\caption{\label{tab:table1}
Parameter sets of the nuclear equation of state.}
\setlength{\tabcolsep}{1.4pt}
\begin{tabular}{|l|cccccc|}
\hline
EoS & $K_0$(MeV) & $\alpha$(MeV) & $\beta$(MeV) & $\gamma$ & $t_{md}$(MeV) & $a_{md}$($\frac{c^2}{GeV^2}$) \\ \hline
SM & 200 & -393 & 320 & 1.14 & 1.57 & 500 \\
H & 300 & -165 & 126 & 1.676 & - & - \\
HM & 380 & -138 & 60 & 2.084 & 1.57 & 500 \\ \hline
\end{tabular}
\end{table}

On the event-by-event basis, the cumulants of particle multiplicity distributions can be calculated in the form
\begin{eqnarray}
&C_1=M=\langle N \rangle, \\
&C_2=\sigma^2=\langle (\delta N)^2\rangle, \\
&C_3=S\sigma^3=\langle (\delta N)^3\rangle, \\
&C_4=\kappa \sigma^4=\langle (\delta N)^4\rangle-3\langle (\delta N)^2 \rangle^2 .
\end{eqnarray}
Here $\delta N=N- \langle N \rangle$ with N being the number of particles in a given acceptance (e.g., rapidity or transverse momentum window) for a single event.
$M$ is the mean value of particle multiplicity in the chosen window calculated from averaging all considered events, $\sigma$ is the standard deviation,
$S$ is the skewness which measures the degree of asymmetry of a distribution, and $\kappa$ is the kurtosis which is also a descriptor of the shape of a distribution.
Usually, the ratios of cumulants are used to cancel the unknown volume dependence and directly compared with theoretical calculations,
\begin{eqnarray}
&C_2/C_1=\sigma^2/M,\\
&C_3/C_1=S \sigma^3/M,\\
&C_3/C_2=S \sigma,\\
&C_4/C_2=\kappa \sigma^2.
\end{eqnarray}
According to the Delta-theorem \cite{error estimation}, the statistical error of the cumulants and their ratios can be approximated as follows:
\begin{eqnarray}
&error(C_r) \propto \sigma^r/ \sqrt{n},\\
&error(C_r/C_2) \propto \sigma^{(r-2)}/ \sqrt{n}.
\end{eqnarray}
Here $n$ is the total number of events. In this work, we will focus on Au+Au collision at beam energy of 1.23 GeV$/$nucleon with impact parameter $b$=0 fm.
To do so, more than 2 000 000 events for each cases are simulated to accumulate high statistics on observables. As will be shown later, the statistical errors for the first three moments are quite small, but the errors for $C_4$ and the corresponding kurtosis are still large. One can estimate from Eq.(11) and (12) that, to cut the original error in half would need four times the number of events than before.

\begin{figure}[h!]
\centering
  \includegraphics[width=9.5cm]{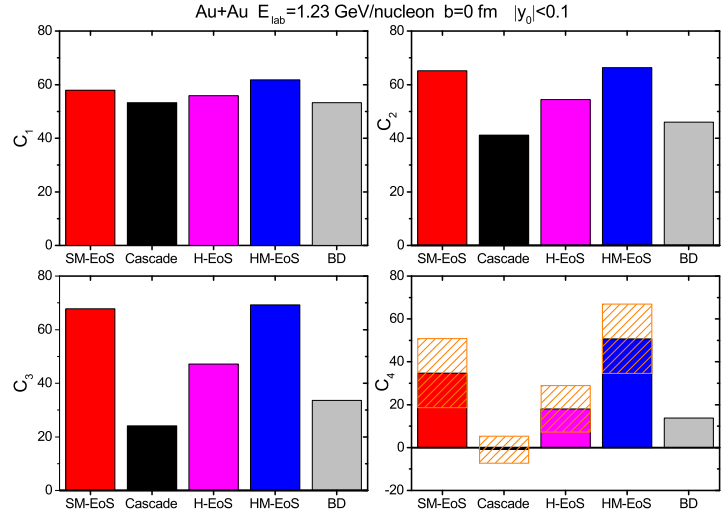}
  \caption{ The moments of all baryons (protons and neutrons) multiplicity distributions produced in the central Au+Au collisions at beam energy of 1.23 GeV/nucleon. The rapidity cut $-0.1<y_0<0.1$ (around mid-rapidity) is considered as an example. The results calculated with SM, H, HM, and cascade mode are compared to the results obtained from the binomial distribution (BD). Error bars whenever not showing, are smaller than symbols size.}
  \label{fig1}
\end{figure}

\begin{figure}[h!]
\centering
  \includegraphics[width=9.5cm]{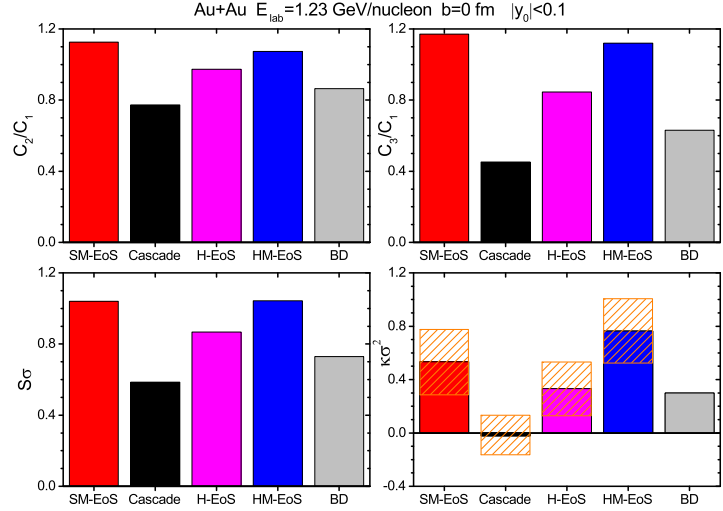}
  \caption{The same as Fig.\ref{fig1} but for the cumulant ratios.}
  \label{fig2}
\end{figure}

Figure \ref{fig1} compares the cumulants of all baryons (protons and neutrons) distribution around mid-rapidity ($-0.1<y_0<0.1$) from different scenarios. The small rapidity cut is chosen to weaken the influences of baryon number conservation. The $C_2$, $C_3$ and $C_4$ for the binomial distribution (BD) are obtained by $C_2$=$Np(1-p)$, $C_3$=$Np(1-2p)(1-p)$, and $C_4$=$Np[1-6p(1-p)](1-p)$ with $N$=394 being the total baryon number and $p$=$C_1$/$N$ being the probability in a given rapidity window, here the $C_1$ is taken from simulation in the cascade mode. We have tested that the result for the BD will not change too much if the $C_1$ value is taken from other scenarios. We take the BD as a baseline because it essentially assumes uncorrelated baryon emission, and the global baryon number conservation has been obeyed. The mean value ($C_1$) obtained with HM is the largest one while that obtained with cascade mode is the smallest one, this is because the strong repulsive interaction in the high-density phase in the case of HM provides more blocking and makes nucleons more likely to be emitted perpendicular to the reaction plane. Both $C_2$ and $C_3$ calculated within the cascade mode are smaller than those calculated with mean field potential, indicating that nuclear interaction enlarges the correlation (i.e., the fluctuation of $\delta N$) between particles. This is due to the attractive nature of the nuclear interaction at sub-normal densities near freeze-out. If one compares the $C_2$ and $C_3$ obtained with H to that obtained with HM and SM modes, it is found that the cumulants are less sensitive to the incompressibility $K_0$ but more sensitive to the momentum-dependent component of the nuclear potential. In addition, it can be seen that the cumulants obtained with mean field potential modes are larger than that obtained with the BD, but the cumulants obtained from the cascade mode are smaller than the BD baseline. This implies the mean field potential and the pure collision term may lead to opposite corrections. In Ref\cite{Ye:2018vbc}, the cumulants as a function of reaction time were presented by both UrQMD model and JAM model, we also found that the cumulants first decrease in the compressed stage (dominated by collision) then increase afterwards (dominated by mean field potential). The cumulant ratios are displayed in Fig.\ref{fig2}. The magnitude of the cumulant ratios varies drastically for the different potential implementations. While the momentum dependent potentials essentially give
the same result, they also show the largest deviation from the BD baseline. Moreover, it can be found again that the cumulant ratios calculated with mean-field potentials (i.e., SM, HM, and H) are larger than the BD baseline while the ones from the cascade mode are smaller than the BD baseline. Although with large error bar, the enhancements for $\kappa \sigma^2$ contributed from the mean-field potential still can be observed and none of the $\kappa \sigma^2$ exceeds the value of 1.

For larger rapidity windows, as can be seen in Ref\cite{Ye:2018vbc}, the differences in the cumulant ratios among different calculations steadily decrease, and their values generally approach the limiting values obtained from the BD for p = 1, i.e., $C_2$/$C_1$ = $C_3$/$C_1$ = 0, and $S\sigma$=1, due to the dominant contribution from baryon number conservation. Besides the effects of mean field potential, the influence of clustering on the cumulants has also been studied in Refs\cite{Ye:2018vbc,Steinheimer:2018rnd}. We found that the cumulant ratios for free baryons and free protons are also enhanced by the mean field potential, as compared to the cascade mode. However, the enhancement of free protons is apparently reduced compared to that for all baryons. These results manifest that the clusterization effect also plays an important role on the cumulants of particle multiplicity distributions at intermediate energies. In addition, we found that the cumulant ratios for free protons are not equivalent to that for for free baryons, as the number of protons is not a half of baryons and the protons (neutrons) cannot completely forget their initial isospin through collisions.

Besides the cumulants in momentum space (with different rapidity window), we also studied the cumulants in the coordinate space (with different length of box around the collision point x=y=z=0) in Ref\cite{Steinheimer:2018rnd}, it is found that coordinate space correlations are not equivalent to the momentum space correlations. Whether the correlations in coordinate space can be completely translated to momentum space correlations which can be measured in experiment remains an open question.

In summary, within the ultrarelativistic quantum molecular dynamics (UrQMD) model, we studied the cumulants and their ratios for baryon multiplicity distributions in the central Au+Au collisions at beam energy of 1.23 GeV$/$nucleon. The influence of mean field potential and clustering on the cumulants was investigated in detail. It was found that both those effects will impact the cumulants and their ratio, and the mean field potential enlarges the cumulants and their ratios as it provides more correlations between particles.

\section{Acknowledgement}
We thank Jan Steinheimer, Yasushi Nara, Hao-jie Xu, Pengcheng Li, Dinghui Lu and Horst Stoecker for helpful discussions in studies on this topic. This work is supported
in part by the National Natural Science Foundation of China (Grants No. 11875125, No. 11847315, and No. 11505057), and the Zhejiang Provincial Natural
Science Foundation of China (Grant No. LY18A050002 and LY19A050001).



\end{document}